\documentclass[%
 reprint,
 amsmath,amssymb,
 aps,
]{revtex4-2}

\usepackage{graphicx}
\usepackage{dcolumn}
\usepackage{bm}
\usepackage{color}

\newcommand{\Sr}{Sr$_2$RuO$_4$}
\newcommand{\al}{\alpha}

\begin{document}


\title{Ultrasound evidence for a two-component superconducting order parameter in Sr$_2$RuO$_4$}

\author{S. Benhabib$^{1,\dagger}$, C. Lupien$^{2,\dagger}$, I. Paul$^{3,*}$, L. Berges$^1$, M. Dion$^2$, M. Nardone$^1$,A. Zitouni$^1$, Z.Q. Mao$^{4,5}$, Y. Maeno$^{4,6}$, A. Georges$^{7,8}$, L. Taillefer$^{2,6,*}$ and C. Proust$^{1,6,*}$}

\affiliation{
$^1$Laboratoire National des Champs Magn\'{e}tiques Intenses (CNRS, EMFL, INSA, UGA, UPS), Toulouse 31400, France\\
$^2$Institut Quantique, D\'{e}partement de physique \& RQMP,
Universit\'{e} de Sherbrooke, Sherbrooke, Qu\'{e}bec, J1K 2R1, Canada \\
$^3$Laboratoire Mat\'{e}riaux et Ph\'{e}nom\`{e}nes Quantiques, Universit\'{e} de Paris, CNRS, F-75013, Paris, France\\
$^4$Department of Physics, Kyoto University, Kyoto 606-8502, Japan\\
$^5$Department of Physics, The Pennsylvania State University, University Park, PA 16803 USA\\
$^6$CIFAR, Toronto, Ontario, M5G 1M1, Canada\\
$^7$Center for Computational Quantum Physics, Flatiron Institute, New York, NY 10010 USA\\
$^8$Coll\`{e}ge de France, 75005 Paris, France\\
}


\begin{abstract}
The quasi-2D metal \Sr~is one of the best characterized unconventional superconductors, yet the nature of
its superconducting order parameter is still highly debated \cite{Maeno94,Rice95,Mackenzie03}. This information is crucial to determine the pairing mechanism of Cooper pairs.
Here we use ultrasound velocity to probe the superconducting state of \Sr. This thermodynamic probe is symmetry-sensitive and can help to identify the superconducting order symmetry \cite{Walker02, Sigrist02}. 
Indeed, we observe a sharp jump in the shear elastic constant
$c_{66}$ as the temperature is raised across the superconducting transition at $T_c$.
This directly implies that the superconducting order parameter is of a two-component nature.
Based on symmetry argument and given the other known properties of \Sr~\cite{Hassinger17, Pustogow19, Li19}, we discuss what states are compatible with this requirement and propose that the two-component order parameter, namely $\lbrace d_{xz}; d_{yz} \rbrace$, is the most likely candidate.

\end{abstract}

\pacs{Valid PACS appear here}

\maketitle


%
In conventional superconductors, the pairing mechanism originates in the interaction of electrons with phonons and the resulting superconducting order parameter has $s$-wave symmetry. The Fermi statistic imposes that the orbital symmetric state of two electrons must combine with an antisymmetric spin singlet part. Contrariwise, in superfluid $^3$He, the ferromagnetic spin fluctuations being the mechanism responsible for pairing, the resulting spin states are triplet. In this case, the orbital part of two $^3$He atoms has $p$-wave symmetry \cite{Legett75}.
For 25 years, superconductivity of \Sr~has been viewed as an electronic analog of superfluid $^3$He \cite{Maeno94,Rice95,Mackenzie03}.
The initial report of the temperature independent spin susceptibility through $T_c$  \cite{Ishida98}
and the indication of time-reversal
symmetry breaking  \cite{Luke98,Xia06} pointed to a spin triplet chiral $p$-wave order parameter. 
%
However, several experiments are in contradiction with this scenario \cite{MacKenzie17}, for example
the lack of edge currents  \cite{Kirtley07}, the Pauli limiting critical field  \cite{Deguchi02}
and the absence of a cusp in the dependence of $T_c$ on uniaxial strain  \cite{Hicks14}.
Importantly, evidence of line nodes in the gap from specific heat \cite{Nishizaki00}, ultrasound attenuation \cite{Lupien01} and thermal conductivity \cite{Suzuki02, Hassinger17} is not compatible with a chiral $p$-wave order parameter, that has no symmetry-imposed node (for a two-dimensional Fermi surface).
Recently, measurements of the NMR Knight shift were carefully revisited and a clear drop
in the spin susceptibility below $T_c$ was detected \cite{Pustogow19},
pointing to  an order parameter with even parity.
As a result, the chiral $p$-wave order parameter is excluded and the nature of the superconducting state in \Sr~is now a wide open question.\\

%

\begin{table}[h]
\centering
\begin{tabular}{|c|c|c|c|}
  \hline
  $\Gamma$ & Basis function & Strain component & Elastic constant   \\
  \hline
  A$_{1g}$ & a($k_x^2$+$k_y^2$)+b$k_z^2$ & $u_{xx}$+$u_{yy}$, $u_{zz}$ & (c$_{11}$+c$_{12}$)/2, c$_{33}$ \\
  A$_{2g}$ & $k_x k_y$($k_x^2$-$k_y^2$) & none & none \\
  B$_{1g}$ & $k_x^2$-$k_y^2$ & $u_{xx}$-$u_{yy}$ & (c$_{11}$-c$_{12}$)/2 \\
  B$_{2g}$ & $k_x k_y$ & $u_{xy}$ & c$_{66}$ \\
  E$_{g}$ & $k_x k_z$, $k_y k_z$ & $u_{xz}$, $u_{yz}$ & c$_{44}$ \\
  \hline
\end{tabular}
\caption{\label{tab1} Irreducible representation of the strain tensor for the D$_{4h}$ point group.}
\end{table}
Sound velocity is a powerful thermodynamic probe for order parameter.
For propagation along high-symmetry directions of the crystal,
the sound velocity is $v_s = \sqrt{\dfrac{c_{ij}}{\rho}}$ where $\rho$ is the density of the material
and $c_{ij}$ are the elastic constants defined as the second derivative of the free energy $F$ with respect to the strain $u_{ij}$. In the framework of Landau-Ginzburg theory of phase transition, the observation of a discontinuity in the elastic constant at the superconducting transition is a consequence of the symmetry allowed coupling term between the order parameter $\Delta$ and the strain $u$, $\lambda \lvert \Delta \rvert^2 u$, where $\lambda$ is a coupling constant \cite{Rehwald73}.
As part of the free energy, this coupling term is invariant under all the operations of
the point group, i.e. it belongs to the A$_{1g}$ representation. Table~\ref{tab1} lists the irreducible strains corresponding to the point group D$_{4h}$ for the tetragonal symmetry of \Sr~(see the corresponding product table in the SI section 6). If the superconducting order parameter is one-component, then $\lvert \Delta \rvert^2$ belongs to the A$_{1g}$ representation.
Consequently, the strain variable $u$ can only belong to the A$_{1g}$ representation, i.e. it corresponds to a longitudinal sound wave. A jump in the longitudinal elastic constant is observed at $T_c$ in many superconductors and is directly related to the jump in the specific heat at $T_c$ and the strain dependence of $T_c$ via the Ehrenfest relation \cite{Sigrist02}.
If an unusual jump in the elastic constant associated
with a shear mode (B$_{1g}$ or B$_{2g}$ representation) is detected at $T_c$, then it necessarily implies that the superconducting order parameter is multi-dimensional \cite{Walker02, Sigrist02, Contreras16}. 

Based on these symmetry arguments,  further developed in this paper, we have performed measurements of longitudinal and transverse sound velocities in \Sr~across the superconducting transition down to 40~mK. The initial measurements \cite{Lupien02} have been confirmed only recently using a different spectrometer (see SI section 2) and by a complementary technique \cite{Ghosh20}.


\begin{table}[h]
\centering
\begin{tabular}{|c|c|c|c|c|}
  \hline
  Elastic & \textbf{k} & \textbf{p} & Sound velocity & Value   \\
  constant&&&(km/s) & (GPa)\\
  \hline
  c$_{11}$ & [100] & [100] & 6.28 & 233 \\
  c$_{44}$ & [100] & [001] & 3.41 & 68.2 \\
  c$_{66}$ & [100] & [010] & 3.3 & 64.3 \\
  (c$_{11}$-c$_{12}$)/2 & [110] & [1$\overline{1}$0] & 2.94 & 51 \\
  \hline
\end{tabular}
\caption{\label{tab3}Definition of the different sound modes measured at $T$ = 4~K. \textbf{k} and \textbf{p} stand for the propagation
and polarization direction, respectively. Sound velocities were obtained at low temperature using the echo spacing.}
\end{table}

\begin{figure}[h]
\centering
\includegraphics[width=0.49\textwidth]{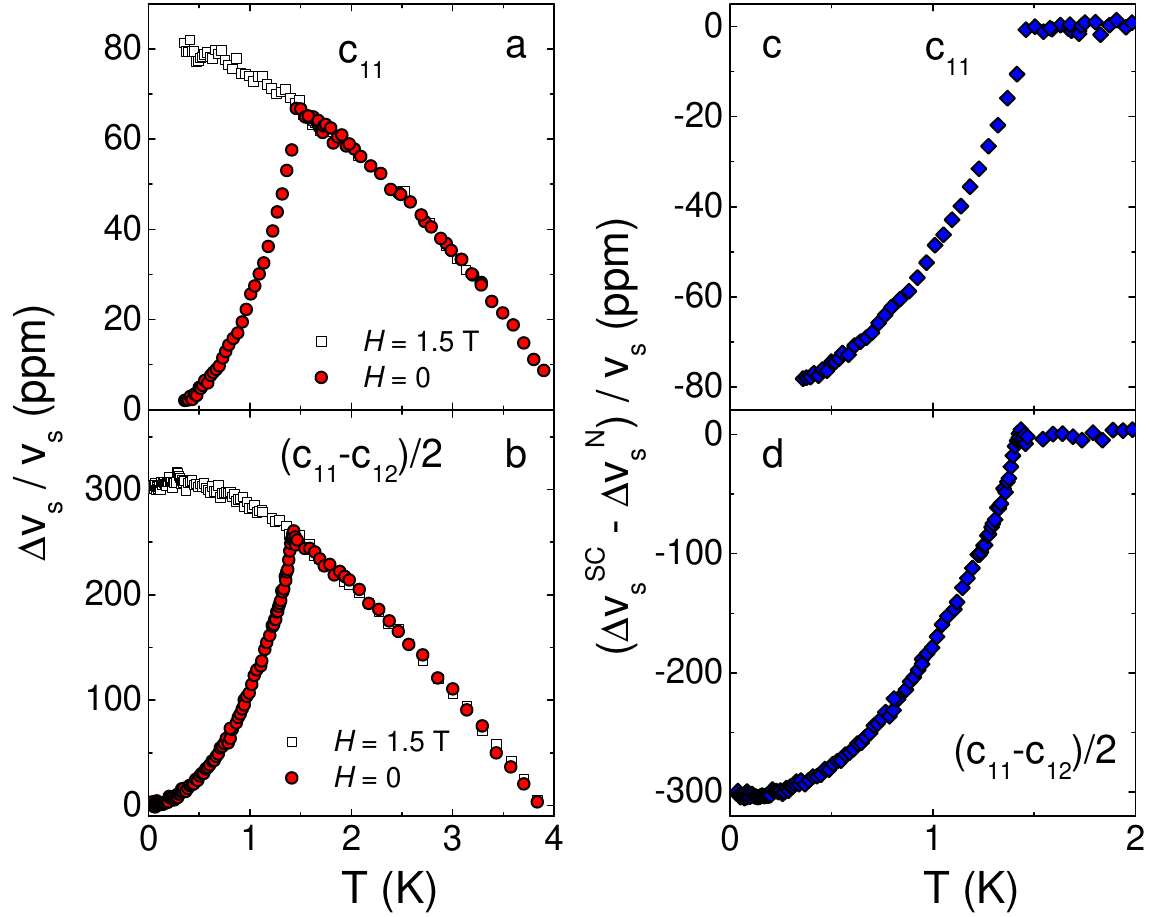}
\caption{
\textbf{Relative change in the sound velocity of \Sr~through $T_c$.}
Temperature dependence of the sound velocity for a) the longitudinal mode $c_{11}$ measured at $f$ = 83 MHz,
b) the transverse mode ($c_{11}$-$c_{12}$)/2 measured at $f$ = 21.5 MHz.
The normal state data (open squares) are obtained by applying
a magnetic field of 1.5~T in the plane, larger than $H_{c2}$.
The superconducting state data (red circles) are measured without
any applied field.
c) Difference between the superconducting state and the normal state,
for the $c_{11}$ mode.
d) Same, for the ($c_{11}$-$c_{12}$)/2 mode.
}
\label{Fig1}
\end{figure}

Table~\ref{tab3} shows the different acoustic modes with the directions of sound propagation and polarization of the transducer.
The value of the sound velocity is obtained from the echo spacing at low temperature ($T$ = 4~K) and can be converted to elastic
constants using $\rho$=5.95~g/cm$^3$. They are in good agreement with resonant ultrasound spectroscopy
measurements \cite{Paglione02,Barber19,Ghosh20}.
Fig.~\ref{Fig1}a and Fig.~\ref{Fig1}b show the temperature dependence of the sound velocity
for the longitudinal mode $c_{11}$ and the transverse mode ($c_{11}$-$c_{12}$)/2, respectively.
Red circles (open squares) correspond to measurements in the superconducting (normal) state. Fig.~\ref{Fig1}c and Fig.~\ref{Fig1}d show the difference between the superconducting state and the normal state for the two modes.
A discontinuity is expected at $T_c$ for the longitudinal mode, $c_{11}$.
We estimate the magnitude of this drop to be $\Delta c_{11} / c_{11} \approx 2$~ppm. This rough estimation is based on the Ehrenfest relation, that links the jump of the sound velocity with the jump of the specific heat and the strain dependence of $T_c$ (see SI section 3).
This small discontinuity at $T_c$ is thus hidden by the strong softening of the longitudinal constant in the superconducting state ($\approx$ 80~ppm between $T_c$ and $T \rightarrow 0$).
A similar, but even stronger softening is observed for the transverse mode ($c_{11}$-$c_{12}$)/2, below $T_c$ (Fig.~\ref{Fig1}b).
These results are qualitatively in agreement with previous measurements \cite{Matsui01, Okuda02}
but the absolute value of their elastic constants differs from ours \cite{Suzuki}.

\begin{figure}[h]
\centering
\includegraphics[width=0.47\textwidth]{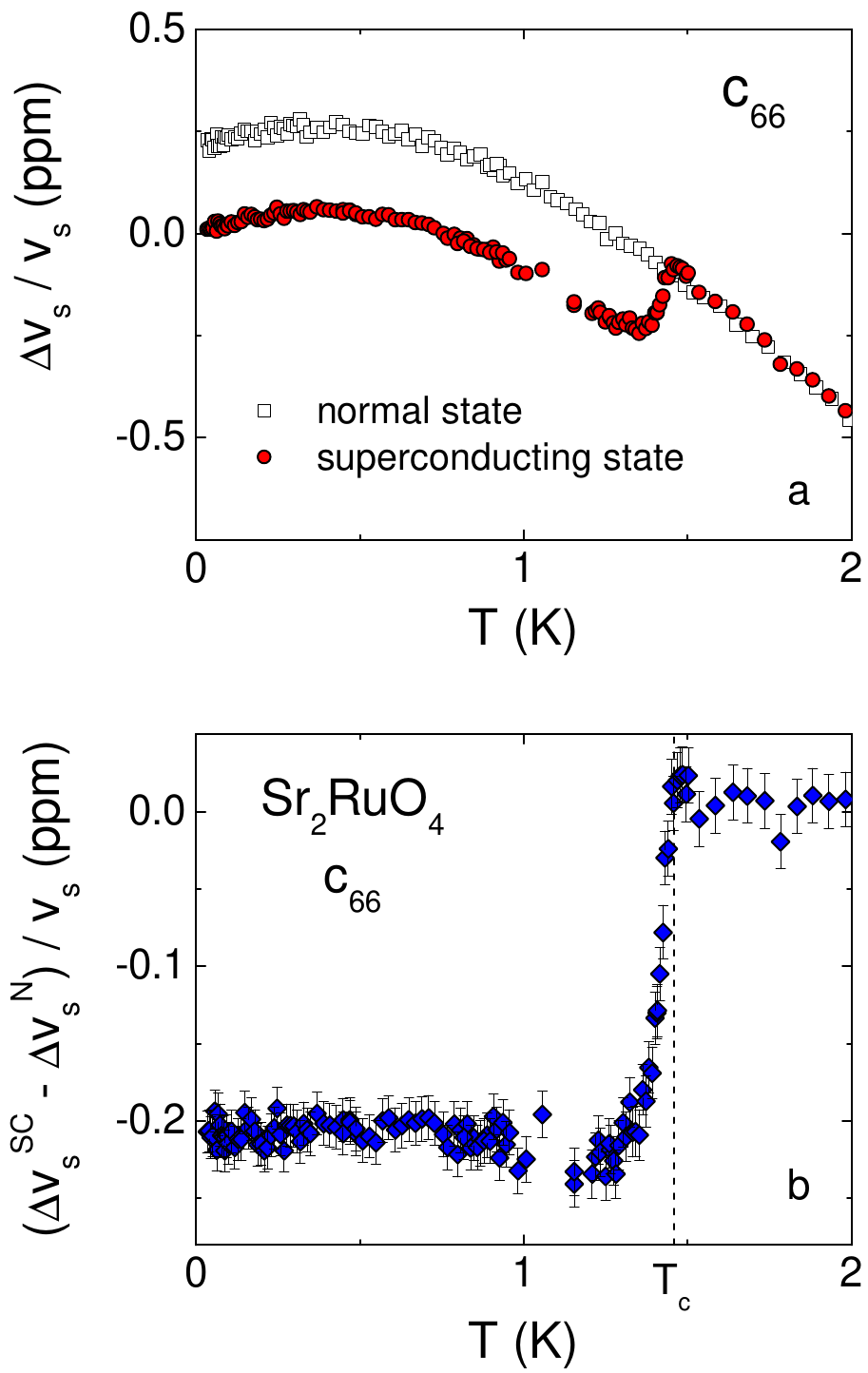}
\caption{
\textbf{Jump in the $c_{66}$ shear modulus at $T_c$.} 
a)
Relative change in sound velocity for the transverse mode $c_{66}$ measured at $f$ = 169 MHz.
The normal state data (open squares) are obtained by applying a field of 1.5 T in the plane.
The superconducting state data (red circles) are measured without any applied field.
b)
Difference between the superconducting state and the normal state, for the $c_{66}$ mode.
A clear discontinuous jump is observed at $T_c$.
The error bar are estimated from a constant voltage amplitude noise on the echoes measured by the phase comparator.
}
\label{Fig2}
\end{figure}

Fig.~\ref{Fig2}a shows the temperature dependence of the sound velocity for the transverse mode, $c_{66}$.
The measurements in the superconducting state ($H$ = 0, red circles) display a sharp discontinuity at the superconducting transition.
The difference in the shear sound velocity between the normal and superconducting states (Fig.~\ref{Fig2}b) shows a small but very clear jump at $T_c$, of magnitude $\simeq 0.2$~ppm, 10 times larger than our experimental resolution.
The exceptional sensitivity of our experiment is due to the very small attenuation of the $c_{66}$ mode \cite{Lupien01}, which enabled us to detect up to $\sim$ 60 echoes (see Fig. S1 in the SI) and to perform a fit on all of them.

Note that if there is any mixing of acoustic modes in the measurement, the small jump in $c_{66}$ can easily be swamped by the huge softening of other modes.
In a second experiment, using a different spectrometer, we were able to again detect the sharp drop below $T_c$ in \Sr, but with some contamination from other modes (see Fig. S2 in the SI).
In the data of Fig.~\ref{Fig2}b, the complete lack of any temperature dependence below 1.3~K down to 0.04~K is strong evidence against such contamination.

This is the key finding of our study: we observe a sharp discontinuity at $T_c$ in the sound velocity for the transverse mode $c_{66}$.
This immediately provides unambiguous evidence that the superconducting order parameter must be of a two-component nature, consistent only with the E$_g$ singlet representation, the E$_u$ triplet representation or an accidentally degenerate combination of two one-dimensional representations.\\
A discontinuity of the sound velocity in the $c_{66}$ mode at $T_c$ of magnitude $\approx$10~ppm was recently detected in \Sr~also by resonant ultrasound spectroscopy performed at $f\sim$2~MHz \cite{Ghosh20}.  The difference in the magnitude of the jump may come from a finite frequency effect on the dynamic of the order parameter. The consequence is a decrease of the amplitude of the anomaly as the frequency increases (see SI section 4 and Ref.~\cite{Sigrist02}). Note that a rough estimation of the magnitude of the $c_{66}$ drop using the Ehrenfest relation is about 2~ppm (see SI section 3).


Let's now turn on the Landau theory describing the strain-order parameter coupling. ${\rm Sr}_2{\rm RuO}_4$ has $D_{4h}$ symmetry, and only the irreducible representation $E$
of the point group is multi-dimensional. In this representation
the superconducting order parameter is a two-component complex variable $(\Delta_A, \Delta_B)$.
If the order parameter has $E_u$ symmetry,
$(\Delta_A, \Delta_B)$ transform as $(x,y)$,
and if the order parameter has $E_g$ symmetry,
they transform as $(xz,yz)$.
For both cases, the Landau-Ginzburg
free energy describing $\Delta$ and the uniform strains $u$ is given by
\begin{align}
\label{eq:free-total}
F = F_{\Delta} + F_u + F_{\Delta-u}.
\end{align}
The superconducting part, expanded to fourth order, is
\begin{align}
F_{\Delta} &= a \left( |\Delta_A|^2 + |\Delta_B|^2 \right)
+ \beta_1^0 \left(|\Delta_A|^2 + |\Delta_B|^2 \right)^2
\nonumber \\
&+ \frac{\beta_2^0}{2} \left[(\Delta_A^{\ast})^2 \Delta_B^2 + {\rm c.c.} \right]
+ \beta_3^0 |\Delta_A|^2 |\Delta_B|^2.
\nonumber
\end{align}
The relevant elastic energy of the uniform strains is
\begin{align}
F_u &= \frac{1}{2} c_{11} \left( u_{xx}^2 + u_{yy}^2 \right) + c_{12} u_{xx} u_{yy} + 2 c_{66} u_{xy}^2
\nonumber \\
&+ \frac{1}{2} c_{33} u_{zz}^2 + c_{13} (u_{xx} + u_{yy})u_{zz},
\nonumber
\end{align}
where $c$'s are the elastic constants in Voigt notation. The cross-coupling term is
\begin{align}
F_{\Delta-u} &= \left[ \alpha_1 (u_{xx} + u_{yy}) + \alpha_2 u_{zz} \right] (|\Delta_A|^2 + |\Delta_B|^2)
\nonumber \\
&+ \alpha_3 (u_{xx} - u_{yy})(|\Delta_A|^2 - |\Delta_B|^2)
\nonumber \\
&+ \alpha_4 u_{xy}(\Delta_A^{\ast} \Delta_B + {\rm c.c.}).
\nonumber
\end{align}

The analysis of the above free energy is standard, and is described in detail in the SI (section 7). Similar expressions for the chiral $p$-wave state have been calculated by other groups \cite{Walker02, Sigrist02, Contreras16}.
Here we quote the main results.

For convenience we define $c_A \equiv (c_{11} + c_{12})/2$ and  $c_O \equiv (c_{11} - c_{12})/2$.
The latter is the orthorhombic elastic constant associated with the shear mode $u_{xx} - u_{yy}$,
while $c_{66}$ is the elastic constant
of the monoclinic shear $u_{xy}$.
Our aim is to calculate the jumps in the shear elastic constants defined by $\delta c \equiv c (T_c^-) - c (T_c^+)$.

The term $F_{\Delta-u}$ renormalizes the fourth order coefficients $\beta_i^0 \rightarrow \beta_i$ with
\begin{align}
\beta_1 &= \beta_1^0 - \frac{1}{2} \left[ \frac{\al_3^2}{c_O} +
\frac{\al_1^2 c_{33} + \al_2^2 c_A - 2 \al_1 \al_2 c_{13}}{c_A c_{33} - c_{13}^2} \right],
\nonumber \\
\beta_2 &= \beta_2^0 - \al_4^2/(4c_{66})
\nonumber \\
\beta_3 &= \beta_3^0 - \al_4^2/(4c_{66})+ 2\al_3^3/c_O.
\nonumber
\end{align}

For the stability of the system,
we need $\beta_1 >0$, and $4 \beta_1 \pm \beta_2 + \beta_3 > 0$. Within these ranges
the following three superconducting phases are possible.

\emph{Case (1) Time reversal symmetry broken superconductor}:
In the region $\beta_2 > (0, \beta_3)$,
we get the time reversal symmetry broken state with $(\Delta_A, \Delta_B) = \Delta_0(1, \pm i)$.
In this phase, there is no spontaneous shear strain, and the tetragonal symmetry is preserved.
The shear moduli jumps are
\begin{subequations}
\begin{eqnarray}
\delta c_{66} &=& \frac{- \al_4^2}{4\beta_2 + \al_4^2/c_{66}},\\
\delta c_O &=& \frac{- 2 \al_3^2}{\beta_2 - \beta_3 + 2\al_3^2/c_O}.
\end{eqnarray}
\end{subequations}
\emph{Case (2) Nematic-monoclinic superconductor}:
In the region $\beta_2 < (0, - \beta_3)$,
we get a nematic solution, $(\Delta_A, \Delta_B) = \Delta_0(1, \pm 1)$,
which breaks the tetragonal symmetry by making the two in-plane diagonal directions inequivalent.
It is accompanied by a spontaneous monoclinic strain, {\it i.e.} $u_{xy} \neq 0$.
The shear moduli jumps are
\begin{subequations}
\begin{eqnarray}
\delta c_{66} &=&  \frac{- \al_4^2/2}{4 \beta_1 + \beta_2 + \beta_3 + \al_4^2/(2c_{66})},\\
\delta c_O &=& \frac{- 2 \al_3^2}{|\beta_2| - \beta_3 + 2\al_3^2/c_O}.
\end{eqnarray}
\end{subequations}
\emph{Case (3) Nematic-orthorhombic superconductor}:
In the region $\beta_3 > (0, \left| \beta_2 \right|)$,
we also get a nematic solution, $(\Delta_A, \Delta_B) = \Delta_0(0, 1)$, or equivalently $\Delta_0(1, 0)$,
which also breaks the tetragonal symmetry by making the two in-plane crystallographic axes inequivalent.
It is accompanied by a spontaneous orthorhombic strain, {\it i.e.} $u_{xx} - u_{yy} \neq 0$.
The shear moduli jumps are
\begin{subequations}
\begin{eqnarray}
\delta c_{66} &=&  \frac{- \al_4^2/2}{\beta_2 + \beta_3 + \al_4^2/(2c_{66})},\\
\delta c_O &=& \frac{- \al_3^2}{2 \beta_1 + \al_3^2/c_O}.
\end{eqnarray}
\end{subequations}
Thus, in all three states the two shear elastic constants,
$c_{66}$ and $c_O$, jump at $T_c$.
In our data, there is a clear jump in $c_{66}$.
However, a jump in $c_O$ could not be resolved,
most likely because of the strong temperature dependence of $c_O(T)$
below the transition.\\


The observed jump in $c_{66}$ at $T_c$
implies that the superconducting order parameter of \Sr~is of a two-component nature.
We now discuss the various implications of this new constraint, in the context of the other known properties of \Sr.

(i) \emph{Discrete symmetry breaking}:
In a two-component scenario, the $U(1)$ symmetry breaking superconducting transition is necessarily
accompanied by a simultaneous discrete symmetry breaking.
For case (1), this discrete symmetry is time reversal leading to a spontaneous magnetization that can be detected in a $\mu$SR measurement, for example. A non-zero $\mu$SR signal below $T_c$ has indeed been reported \cite{Luke98}, but its origin and implications are currently under investigation \cite{Grinenko20}.
For cases (2) and (3), the broken symmetry is tetragonal $D_{4}$ leading to a monoclinic or an orthorhombic  distortion of the tetragonal unit cell, respectively. In principle, it be detected through x-ray diffraction but  no such distortion has been reported yet.

(ii) \emph{Response to uniaxial pressure}:
$T_c$ as a function of the $B_{1g}$ shear strain $u_{xx} - u_{yy}$ is expected to increase linearly.
However experimentally, $T_c$ increases quadratically with the uniaxial strain $\epsilon_{100}$ along [100], and therefore, the cusp at zero strain has not been observed. Note that, due to Poisson effect, $\epsilon_{100}$ is a combination of the $B_{1g}$ shear $u_{xx} - u_{yy}$, and the in-plane $A_{1g}$ longitudinal strain $u_{xx} + u_{yy}$. This implies that at quadratic order in $\epsilon_{100}$, there is a $B_{1g}$ perturbation that should lead to a splitting of the superconducting transition, if the order parameter is the $(1, i)$ or $(1,1)$ type (see SI section 8). But for $(1,0)$, one expects a single transition, with enhanced $T_c$. 
Since specific heat measurements detect no splitting of the superconducting transition under the application of strain (at least along the [100] direction) \cite{Li19}, the behavior of \Sr~under uniaxial pressure argues in favor of the $(1, 0)$ order parameter. (Here we assume that the strain-independent transition observed recently by zero-field muon spin relaxation \cite{Grinenko20} is not related to the superconducting state).

(iii) \emph{Spin-wavevector content of Cooper pairs}:
The drop of the Knight shift below $T_c$ is strongly suggestive of an even parity order parameter \cite{Pustogow19, Ishida19}. Assuming only intraband pairing, for singlets, the lowest harmonic is a $d$-wave solution $(\Delta_A, \Delta_B) = \Delta_0(k_xk_z, k_y k_z)$. For triplets, in order to be consistent with recent NMR as well as polarized neutron scattering data \cite{Petsch20}, the lowest harmonic triplet order parameter is the $p$-wave solution $(\Delta_A, \Delta_B) = \Delta_0 k_z (\hat{d}_x, \hat{d}_y)$.

(iv) \emph{Line nodes}:
Experimentally, thermal conductivity measurements show that the gap has vertical line nodes \cite{Hassinger17}. Whether the gap can also have horizontal line nodes is a quantitative question (see SI section 5).
Recent data from quasiparticle interference (QPI) experiments are interpreted in terms of vertical line nodes along the diagonal\cite{Sharma19}, as in a $d_{x^2-y^2}$ state.
Data on the variation of specific heat as a function of the angle of an in-plane magnetic field relative to the crystal have been interpreted either in terms of vertical line nodes of the $k_xk_z$-type \cite{Deguchi04} (i.e. rotated by 45 $\deg$ compared to the $d_{x^2-y^2}$ state) or horizontal line nodes \cite{Kittaka18}.
All states that we discuss within the E$_g$ and E$_u$ representations necessarily have horizontal line nodes.
In the singlet sector (E$_g$), the states $(1,0)$ and $(1,1)$, which break tetragonal symmetry,
also have vertical line nodes (respectively in the [100] and [110] directions). However, the state $(1,i)$, which breaks time reversal symmetry, typically does
not have vertical line nodes, unless the pairing leads to a Bogoliubov Fermi surface \cite{Suh19}.
In the triplet sector (E$_u$), the $p$-wave solutions do not have vertical line nodes.
In order to have triplet solutions also consistent with vertical line nodes
one needs to consider a higher harmonic $f$-wave solution
$(\Delta_A, \Delta_B) = \Delta_0k_z(k_x^2-k_y^2)( \hat{d}_x, \hat{d}_y)$.

(v) \emph{Accidental degeneracy}:
Until now, we only considered the two-dimensional irreducible representation $E$. In principle, one can also get
a two-dimensional $\Delta$ if two one-dimensional representations become accidentally degenerate. The advantage of
such a scenario is that it allows the possibility of having a finite jump in $c_{66}$ while having no jump of $c_O$.
Thus the $(s+d)$-wave solution $(\Delta_A, \Delta_B) = (1, k_x k_y)$ will have the same free energy structure
as in Eq.~\eqref{eq:free-total}, except with $\al_3 =0$. However, such a state is not guaranteed to have line nodes.
Vertical line nodes are present for accidental degeneracy of higher order harmonics such as
the $(d+g)$-wave solution $(\Delta_A, \Delta_B) = (k_x^2 - k_y^2)(1, k_x k_y)$ \cite{Kivelson20}.\\
%

If we put together the following arguments: 
1) The superconducting order parameter has two components;
2) the gap has vertical line nodes (from thermal conductivity \cite{Hassinger17});
3) the order parameter has even parity (from NMR \cite{Pustogow19});
4) uniaxial strain does not split the superconducting transition (from specific heat under strain \cite{Li19}).
And if we disregard the evidence of time reversal symmetry breaking \cite{Luke98, Xia06, Grinenko20} then the only possible candidate is the (1,0) state in E$_g$ representation, namely the nematic state $k_xk_z$ (or $k_yk_z$), with both horizontal and vertical line nodes. The onset of this nematic state will be accompanied by an orthorhombic distortion of the lattice, and the formation of nematic domains. These structural changes should be detectable by x-ray diffraction measurements. \\


\clearpage
\section*{Methods}
\subsection*{Samples}
Our experiments were carried out on two oriented pieces cut from a single high-quality crystal of \Sr~grown
by the travelling-solvent floating-zone technique\cite{Mao00}.
$T_c$ defined as the peak of the magnetic susceptibility is 1.4~K. Samples were polished to 1 $\mu$m roughness, with two opposite faces whose parallelism was estimated to be better than 1.5 $\mu$m/mm. The alignment of the polished faces relative to the crystal axes was determined by Laue back reflection to be less than 0.5 $\deg$ off axis for the measurements shown in the main. One sample was aligned with the polished face perpendicular to (100) and the other sample was perpendicular to (110).

\subsection*{Ultrasound measurements}
The measurements were performed with a pulse-echo technique using two home built spectrometers and commercial LiNbO$_3$ transducers.
Most of the measurements were performed in a dilution refrigerator in Toronto~\cite{Lupien02}
(by C.~Lupien and C.~Proust, in the Taillefer lab),
with the transducer bonded to the crystals with a thin layer of optical coupling compound (Dow Corning 20-057).

The amplitude and the phase were measured using a phase comparator between a reference signal and the signal from the sample. The outputs of the phase comparator, $I$ and $Q$, give the amplitude of the signal $A=\sqrt{I^2+Q^2}$ and the phase $\phi = {\rm arctan}(Q/I)$.

The $c_{66}$ mode
was later re-investigated in a separate experiment, performed in Toulouse,
using another spectrometer, in a $^3$He refrigerator with the transducer bonded to the crystals with AngstromBond glue (AB9110).
In this case, the amplitude of the signal was measured using a logarithmic amplifier.
The phase was measured using a phase comparator but with a limiting amplifier at the input in order to have a perfect decoupling between the phase and the amplitude of the signal.

In the reflective configuration (using only one transducer),
the sound wave propagates along the sample forward and is reflected back by the parallel opposite face, and hits again the same transducer.
Fig.~S1 shows an example of the amplitude of the signal for the $c_{66}$ mode
measured in Toronto.
Each forward-and-back travel of the sound wave corresponds to one echo and up to ~60 echoes were detected.
The sound velocity variation is given by:

\begin{equation}
\label{eq:velocity}
\frac{\Delta v_s}{v_s}=\frac{1}{\omega} \frac{\partial \phi}{\partial t}
\end{equation}
where $\omega=2 \pi f$ and $f$ is the frequency of the measurement. Since the phase is linearly increasing with time, we can perform a linear fit to the phase of different echoes versus time. By using all echoes in a weighted fit, the noise is reduced and the sensitivity of the measurement is close to 0.02 ppm.\\
\\
\textbf{Acknowledgements}
We thank J. Chang, C. Kallin, S. A. Kivelson, V. Madhavan, D. LeBoeuf, B. J. Ramshaw, G. Rikken, M. Sigrist, D. Vignolles and M. B. Walker, for helpful and stimulating discussions.
Part of this work, associated with the PhD thesis of C.L. working with C.P. under the supervision of L.T., was performed at the University of Toronto.
C.P. acknowledges support from the EUR grant NanoX n\textsuperscript{o}ANR-17-EURE-0009 and from the ANR grant NEPTUN n\textsuperscript{o}ANR-19-CE30-0019-01. L.T. acknowledges support from the Canadian Institute for Advanced Research (CIFAR) as a CIFAR Fellow and funding from the Natural Sciences and Engineering Research Council of Canada (NSERC; PIN: 123817), the Fonds de recherche du Québec - Nature et Technologies (FRQNT), the Canada Foundation for Innovation (CFI), and a Canada Research Chair. This research was undertaken thanks in part to funding from the Canada First Research Excellence Fund. Y.M. acknowledges support from JSPS Kakenhi (Grants JP15H5852, JP15K21717, and JP17H06136) and the JSPS-EPSRC Core-to-Core Program “Oxide-Superspin (OSS)”.\\
\\
\textbf{Author Contributions}
C.L. and C.P. performed the ultrasound measurements in Toronto. S.B., L.B. and C.P. performed the ultrasound measurements in Toulouse. S.B., C.L., L.B., M.D. and C.P. analysed the data. I.P. performed the calculations, with the input of A.G. M.N. and A.Z. conceived and realized the $^3$He cryostat in Toulouse. Z.Q.M and Y.M. prepared and characterized the \Sr~sample. I.P., L.T and C.P wrote the manuscript, in consultation with all authors. I.P., L.T. and C.P. co-supervised the project.\\
\\
\textbf{Competing interests}
The authors declare no competing interests.\\
\\
\textbf{Data availability}
All data that support the findings of this study are available from the corresponding authors on request.\\
\\
$\dagger$ These authors contributed equally to this work.\\
\newline
$*$ Correspondence and requests for materials should be addressed to I.P (indranil.paul@univ-paris-diderot.fr), L.T. (louis.taillefer@usherbrooke.ca) or C.P. (cyril.proust@lncmi.cnrs.fr).
%
%


\begin{references}
\bibitem{Suzuki}
Note that T. Suzuki and I. Ishii remeasured and confirmed recently the values of the elastic constants that are similar to the ones presented in the present paper.
\end{references}
\end{document}